\documentclass[aps,twocolumn,pra,10pt,superscriptaddress,showpacs,floatfix]{revtex4-2}
\usepackage{amsmath}
\usepackage{amssymb}
\usepackage{graphicx}
\usepackage{appendix}
\usepackage{color}
\usepackage{url}
\usepackage{braket}
\usepackage{booktabs}
\usepackage[usenames,dvipsnames]{xcolor}
\usepackage[colorlinks=true,linkcolor=Blue,urlcolor=Blue,citecolor=Blue]{hyperref}
\usepackage{physics}
\allowdisplaybreaks[2]
\begin{document}
	
	\title{Enhancing ground-state cooling of center-of-mass motions via quantum squeezing from magnon nonlinearity}
	\author{Jiate Xu}
	\author{Xinqian Cui}
	\affiliation{Department of Physics, Hangzhou Dianzi University, Hangzhou 310018, China}
	
	\author{Guolong Li}
	\email{glli@hdu.edu.cn}
	\affiliation{Department of Physics, Hangzhou Dianzi University, Hangzhou 310018, China}
	\affiliation{Zhejiang Key Laboratory of Quantum State Control and Optical Field Manipulation, Hangzhou Dianzi University, Hangzhou 310018, China}
	\date{March 8, 2026}
	
	\begin{abstract}
	Cooling massive oscillators to quantum ground state is an essential prerequisite for their precise control, quantum memory, and quantum ultrasensitive measurement, etc.
	In a cavity-magnomechanical system, the magnon-mechanical coupling, enhanced by microwave cavity driving, can be utilized to cool the center-of-mass motion of a levitated magnetic sphere.
	In this work, we report that the cooling performance can be further improved by exploiting quantum squeezing stemming from magnonic self-Kerr nonlinearity inherent to the ferrimagnetic yttrium-iron-garnet (YIG) sphere.
	By means of suitable pump driving, the Kerr nonlinearity is converted into quantum squeezing, yielding considerable enhancement of the center-of-mass cooling with properly chosen optimal parameters.
	Moreover, we demonstrate that this improvement mechanism for cooling the massive magnetic sphere still works even in the unresolved-sideband regime where the mechanical frequency is smaller than the magnon decay rate.
	Eventually, we quantify the powers of the driving pumps for practical implementation of our scheme in a typical system.
	Our findings may provide a novel way to quantum fundamental researches and technologies.
	\end{abstract}
	\maketitle
		
	\section{Introduction}
	Cooling massive objects to quantum ground state plays a vital role in modern physics and applied technology, including explorations of macroscopic quantum physics \cite{Romero2011, Rahman2019, Sekatski2014, Gonzalez2021}, ultrahigh-precision sensing \cite{Weiss2021, Abbott2016, Prat2017, Mason2019, Cosco2021}, new physics beyond the standard model \cite{Moore2021}, dark matter searching \cite{Monteiro2020, Higgins2024}, gravitational wave detection \cite{Carney2025}, gravitational decoherence \cite{Bassi2017, Blencowe2013}, and toward the classical-quantum boundary \cite{Bertet2001}.
	Hence, levitation devices of massive objects via optical trapping have attracted considerable attention and achieved significant progress \cite{Monteiro2020PRA, Guccione2013, Michimura2017}.
	Nowadays, spherical magnets become potential macroscopic objects to be levitated and motionally cooled \cite{Huillery2020, Timberlake2019, Brien2019, Jiang2020}, confined in several traps \cite{Seberson2020, Rusconi2017, Gieseler2020} or clamped to an ultrahigh-$Q$ mechanical resonator \cite{vinante2011, Fischer2019}.
	Beyond the tiny-sized cooling \cite{Millen2020, Rademacher2020, Delic2020, Leng2021, Streltsov2021}, millimeter-sized spherical magnets can be levitated above a superconductor in free space \cite{Timberlake2019} or within a microwave cavity \cite{Raut2021}.
	
	Since yttrium iron garnet (YIG), a promising magnetic material, can be levitated within a microwave cavity, they jointly form a cavity magnomechanical (CMM) system via the magnetic-dipole interaction between magnon and microwave modes \cite{Huebl2013, Tabuchi2014, Zhang2014, Goryachev2014, Bai2015, zhang2015}. 
	Magnons, the quanta of collective spin excitations in YIG \cite{Zhang2016, Lachance2019, Potts2021, Yuan2022}, become key elements in quantum information science and condensed matter physics \cite{Zhang2016, Zhang2023, Zuo2024, Rameshti2022, Yuan2022} due to their low dissipation \cite{Huebl2013, Tabuchi2014} and excellent tunability \cite{zhang2017, Wang2016}. 
	An intensive coupling between mechanical and magnon modes has recently been proposed for a levitated YIG sphere in a microwave CMM system, notably characterized by being independent of mass \cite{Kani22}.
	This novel mechanism can guarantee the steady cooling rate for center-of-mass motion of a YIG sphere from femto- to millimeter-sized.
	Most notably, the magnon self-Kerr nonlinearity has been found in YIG \cite{Wang2016, Shen2021}, causing mechanical bistability \cite{Shen22} in a variety of physical systems \cite{Chan2001, Sapmaz2003, Badzey2005, Cottone2009, Chen2012, Ricci2017}.
	This kind of magnon nonlinear effect, originating from the intrinsic magnetocrystalline anisotropy in YIG, may be favorable to enhancing cooling of center-of-mass motion in CMM systems \cite{Zoepfl2023, Diaz2024}.
	On the other hand, in bad-cavity optomechanics where the cavity decay rate exceeds the mechanical frequency (i.e., unresolved-sideband regime), the detrimental backaction limits the cooling to a finite phonon occupation, and thus should be suppressed \cite{lau2020}.
	This situation also emerges in CMM systems and should be addressed.
	
	The magnon Kerr nonlinearity has not been considered in center-of-mass cooling in CMM systems yet, but in this work we will fill this research gap via thoroughly leveraging this magnon nonlinearity.
	In result, it significantly enhances the cooling.
	The intrinsic cause of the enhancement lies in quantum squeezing stemming from the Kerr nonlinearity when reasonable pumps drive the magnon mode in YIG sphere.
	After linearizing this Kerr nonlinear system due to the pump fields, we derive the optimal squeezing rate and coupling strength in detail and, thus, acquire the corresponding pumps injected into the YIG sphere for driving the magnon mode.
	Crucially, with the optimal parameters, this improvement strategy is applicable to the system even in the unresolved-sideband regime.
	We eventually supply the powers of the driving pumps in a typical system.
	
	This work is structured as follows.
	In Sec.~\ref{sec:model}, we start with the proposed model with magnon Kerr-nonlinearity.
	In Sec.~\ref{sec:dynamics}, we provide the system dynamics, and derive mean final phonon occupancy characterizing the mechanical cooling. 
	In Sec.~\ref{sec:optimization}, we investigate the optimal parameters and display this enhancement in comparison with the cases without utilizing the nonlinearity.
	Sec.~\ref{sec:power} presents the appropriate powers of the pumps which drives microwave and magnon modes.
	Finally, a brief conclusion is given in Sec~\ref{sec:conclusions}.

	\section{Model}\label{sec:model}
	
	\begin{figure}[btp]
		\includegraphics[width=0.9\columnwidth]{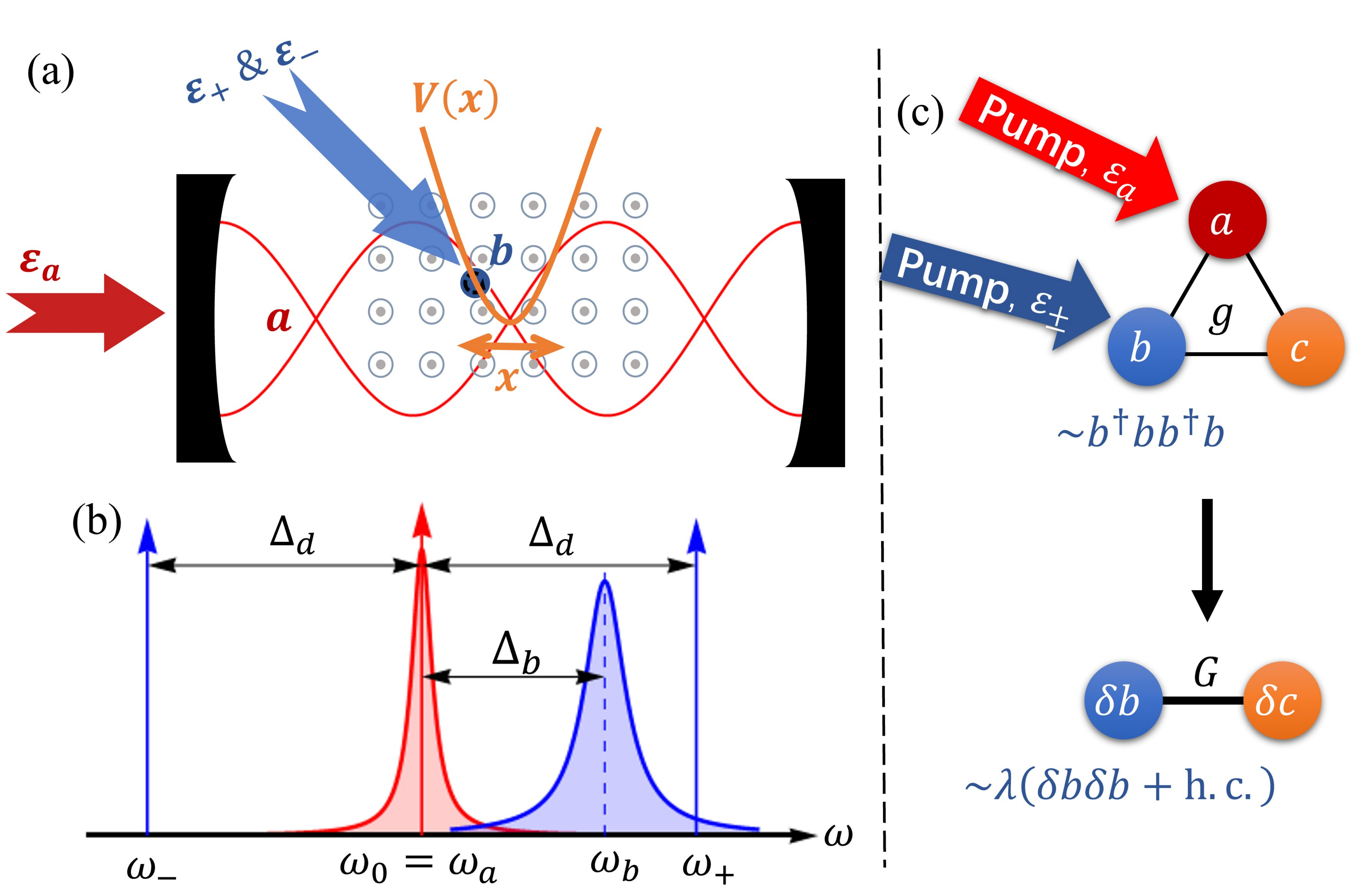}
		\caption{\label{fig:schematic}
			(a) Sketch of the cavity-magnomechanical system.
			A microwave cavity mode $a$ is coupled to a magnon mode $b$ supported by a YIG sphere due to a uniform bias magnetic field.
			The YIG sphere is levitated at a node of the cavity magnetic field and confined in a harmonic potential $V(x)$, moving with center-of-mass displacement $x \propto c + c^\dagger $.
			While a pump with amplitude $\mathcal{E}_a$ drives the cavity mode $a$, a two-tone laser with amplitudes $\mathcal{E}_\pm$ is fed into the YIG sphere to yield the degenerate squeezing of the magnon mode $b$ from its self-Kerr nonlinearity.
			(b) Level scheme for the enhanced cooling.
			The strong pump drives the cavity resonantly, i.e., $\omega_{0} = \omega_{a}$, but has a detuned frequency $\Delta_b$ from the magnon mode.
			The two-tone pump with amplitudes $\mathcal{E}_\pm$ and frequencies $\omega_\pm$ drives the magnon mode, with two equal-spacing detunings $\pm \Delta_d$ from the cavity driving. 
			(c) Under these drivings, the nonlinear tripartite and Kerr processes as Eq.~\eqref{eq:nonlin Ham} can be mapped into a linear Hamiltonian with pump-enhanced coupling strengths $G$ and squeezing coefficient $\lambda$ as Eq.~\eqref{eq:orig lin Hami}.
			As a result, the cavity mode $\delta a$ is decoupled while the magnon and mechanical modes,  $\delta b$ and $\delta c$, are coupled.
		}
	\end{figure}
	
	Fig.~\ref{fig:schematic}(a) depicts the studied CMM framework.
	A YIG sphere is placed at a node of a magnetic field supported by a superconducting MW cavity and, meanwhile, levitated in a harmonic potential which traps its center-of-mass~\cite{Kani22}.
	Such a system consists of a MW cavity mode, a magnon mode supported by the highly polished pure single-crystal YIG, and a mechanical mode of the center-of-mass motion.
	The Hamiltonian is given by 
	\begin{align}\label{eq:org Ham}
		H =\ & \omega_{a} a^\dagger a
		+ \omega_{b} b^\dagger b
		+ \omega_{c} c^\dagger c
		+g \qty(a^\dagger b + a b^\dagger ) \qty(c +c ^\dagger) \notag \\
		&+ \frac{k}{4} b^\dagger b b^\dagger b
		+ i \mathcal{E}_a \qty( a^\dagger e^{-i\omega_{0}t} - a e^{i\omega_{0}t} ) \notag \\
		&+\qty[ i\qty(\mathcal{E}_{+} e^{-i\omega_{+}t} + \mathcal{E}_{-} e^{-i\omega_{-}t}) b^\dagger + \mathrm{h.c.}],
	\end{align}
	where $a$ ($a^\dagger$), $b$ ($b^\dagger$), and $c$ ($c^\dagger$) are the annihilation (creation) operators for the cavity, magnon, and mechanical modes, with frequencies $\omega_{a}$, $\omega_{b}$, and $\omega_{c}$, respectively.
	The creation (annihilation) operator of the magnon mode is defined by the Holstein-Primakoff transformation~\cite{Holstein1940} 
	while the displacement operator $x = x_\mathrm{zpf} (c +c^\dagger )$ represents the mechanical mode of the center-of-mass motion with its zero-point fluctuation $x_\mathrm{zpf}$.
	The fourth term describes the tripartite coupling, with strength $g$, of all these modes, yielded by positioning the trap at a node of the cavity field \cite{Kani22}.
	The fifth them represents magnon self-Kerr nonlinearity with the coefficient $k$ \cite{Shen22}.
	The last two terms show that, while the cavity mode is driven by a pump with strength $\mathcal{E}_a$ and frequency $\omega_{0}$, the magnon mode is specifically driven by two-tone strong field with strength $\mathcal{E}_\pm$ and frequency $\omega_\pm$.
	
	A similar coupling of center-of-mass motion to nitrogen-vacancy (NV) center was realized experimentally via a levitated magnetic particle \cite{Streltsov2021} though, to our knowledge, a direct experimental validation of the specific coupling in Hamiltonian \eqref{eq:org Ham} has not yet been reported.
	That work established the coupling by modulating the displacement of a massive particle, and introduced measurement-based cooling of the center-of-mass motion by using of two-level system (TLS) sensor, such as NV center, rather than sideband cooling with the help of microwave cavity in this work.
	Likewise, another experiment built a coupling of center-of-mass motion to SQUID via superconducting lead-tin sphere \cite{Hofer23}.
	Moreover, in a similar way, the tripartite coupling among single spins, magnons, and displacements have been demonstrated in hybrid systems \cite{Hei23}.
	Owing to the valid theoretical foundation, this system described by Hamiltonian \eqref{eq:org Ham} has been considered as a promising platform to generate entanglement \cite{Kumar25, Zhan25, Bayati24}.
	
	After a frame rotating at the frequency $\omega_{0}$ of the driving field for cavity mode, the Hamiltonian becomes
	\begin{align}\label{eq:nonlin Ham}
		H =\ &\Delta_{a} a^\dagger a
		+ \Delta_{b} b^\dagger b
		+ \omega_{c} c^\dagger c 
		+g \qty(a^\dagger b 
		+ a b^\dagger ) \qty(c +c ^\dagger) \notag \\
		&+ \frac{k}{4} b^\dagger b b^\dagger b
		+ i \mathcal{E}_a \qty( a^\dagger - a ) \notag \\
		&+ i\qty[\qty(\mathcal{E}_{+} e^{-i\Delta_{d}t} + \mathcal{E}_{-} e^{i\Delta_{d}t}) b^\dagger - \mathrm{h.c.}],
	\end{align}
	with a condition as
	$\omega_{+} + \omega_{-} = 2 \omega_{0}$
	and two definitions as
	$\Delta_{a(b)} \equiv \omega_{a(b)} - \omega_{0}$ 
	and
	$\Delta_{d} \equiv \omega_{+} - \omega_{0} = - \qty(\omega_{-} - \omega_{0}) = \qty(\omega_{+} - \omega_{-})/2$.
	We assume $\Delta_{a} = 0$ and the frequency domain of the scheme is shown in Fig.~\ref{fig:schematic}(b).
	The Langivan equations for the above Hamiltonian are
	\begin{align}
	\dot{a} =& -\qty(i \Delta_{a} + \frac{\kappa_a}{2}) a
	- ig b\qty(c + c^\dagger) + \mathcal{E}_a
	+ \sqrt{\kappa_a} a_\mathrm{in}, \notag \\
	\dot{b} =& -\qty(i \Delta_{b} + \frac{\kappa_b}{2}) b
	- ig a\qty(c + c^\dagger) 
	- \frac{i}{2} k b^\dagger b^2 \notag \\
	&+ \qty(\mathcal{E}_{+} e^{-i\Delta_{d}t} + \mathcal{E}_{-} e^{i\Delta_{d}t}) + \sqrt{\kappa_b} b_\mathrm{in}, \notag \\
	\dot{c} =& -\qty(i\omega_{c} +\frac{\kappa_c}{2}) c
	-i g \qty(a^\dagger b + a b^\dagger)
	+\sqrt{\kappa_c} c_\mathrm{in},
	\end{align}
	with decay rate $\kappa_j$ for mode $j$ ($j = a,b,c$).
	
	By splitting all modes ($a$, $b$, and $c$) into the steady-state values ($A_{0,\pm}$, $B_{0,\pm}$, and $C_{0,\pm}$) and the fluctuation operators ($\delta a$, $\delta b$, and $\delta c$)  as
	\begin{align}
	a = & A_0 + A_{-} e^{i\Delta_{d}t} + A_{+} e^{-i\Delta_{d}t} + \delta a, \notag \\
	b = & B_0 + B_{-} e^{i\Delta_{d}t} + B_{+} e^{-i\Delta_{d}t} + \delta b, \notag \\
	c = & C_0 + C_{-} e^{i\Delta_{d}t} + C_{+} e^{-i\Delta_{d}t} + \delta c,
	\end{align}
	above Langivan equations indicate that the steady-state values are satisfied with
	\begin{align}\label{eq:steady eqs}
	\mathcal{E}_a =& \qty(i\Delta_{a} + \frac{\kappa_a}{2}) A_0 \notag \\
	&+ i g \qty[B_{-} \qty(C_{+} + C_{-}^\ast) + B_{+} (C_{-} + C_{+}^\ast)], \notag \\
	\mp i \Delta_{d} B_\pm =& -\qty(i\Delta_{b} + \frac{\kappa_b}{2}) B_\pm
	- i g A_0 \qty(C_\pm + C_\mp^\ast) \notag \\
	&- \frac{i}{2} k B_\pm \qty(|B_\pm|^2 + 2|B_\mp|^2) + \mathcal{E}_\pm, \notag \\
	\mp i \Delta_{d} C_\pm =& - \qty(i\omega_{c} + \frac{\kappa_c}{2}) C_\pm
	- i g (A_0^\ast B_\pm + A_0 B_\mp^\ast), \notag \\
	A_\pm =& B_0 = C_0 =0, 
	\end{align}
	and the fluctuations for every modes obey the following linearized equations,
	\begin{align}\label{eq:linear lang eqs}
	\delta \dot{a} =& - \qty(i\Delta_{a} + \frac{\kappa_a}{2}) \delta a
	+\sqrt{\kappa_a} a_\mathrm{in}, \notag \\
	\delta \dot{b} =& - \qty(i\Delta'_{b} + \frac{\kappa_b}{2}) \delta b
	- i G \qty(\delta c + \delta c^\dagger)
	- i \lambda e^{i\theta} \delta b^\dagger
	+ \sqrt{\kappa_b} b_\mathrm{in}, \notag \\
	\delta \dot{c} =& -\qty(i\omega_{c} + \frac{\kappa_c}{2}) \delta c
	- i G \qty(\delta b + \delta b^\dagger)
	+ \sqrt{\kappa_c} c_\mathrm{in}.
	\end{align}
	Here we adjust the detuning frequency as
	$\Delta'_{b} = \Delta_{b} + k \qty(|B_{-}|^2 + |B_{+}|^2)$,
	and respectively denote the driving-enhanced magnon-mechanical coupling strength and degenerate magnon squeezing coefficient as
	\begin{equation}\label{eq:enhanced factor}
		G = g A_0, \quad \lambda e^{i\theta} = k B_{+} B_{-}.
	\end{equation}
	Without loss of generality, we take $A_0$ as a real number which yields the real coupling strength $G$.
	Moreover, the complex-valued solutions $B_\pm$ jointly determine the squeezing coefficient with modulus $\lambda$ and phase $\theta$.
	
	We drop the prefix $\delta$ of the fluctuations in the following  for convenience.
	Based on Eqs.~\eqref{eq:linear lang eqs}, the cavity mode is decoupled to other modes, and
	the corresponding linearized Hamiltonian is given by 
	\begin{align}\label{eq:orig lin Hami}
	H_\mathrm{lin}
	=& \ \Delta'_{b}  b^\dagger b
	+ \omega_{c} c^\dagger c
	+ G ( b + b^\dagger)
	( c + c^\dagger)  \notag \\
	&+ \frac{\lambda}{2} \qty( b^2 e^{-i\theta} +  b^{\dagger2} e^{i\theta}),
	\end{align}
	as shown in Fig.~\ref{fig:schematic}(c).
	We further introduce the Bogoliubov transformation as
	\begin{align}\label{eq:bogo transf}
	\beta = b e^{i\phi} \cosh r
	+ b^\dagger e^{i(\phi + \theta)} \sinh r,
	\end{align}
	where the coefficient $r$ is set as
	\begin{align}\label{eq:r}
	\sinh 2r = \frac{\lambda}{\sqrt{\Delta_{b}'^2 - \lambda^2}}, \quad
	\cosh 2r = \frac{\Delta_{b}'}{\sqrt{\Delta_{b}'^2 - \lambda^2}},
	\end{align}
	and the phase $\phi$ is chosen to satisfy the real number condition as
	\begin{equation}\label{eq:miu}
	\mu = e^{-i\phi} \qty(\cosh r - e^{-i\theta} \sinh r) \in \mathbb{R}.
	\end{equation}
	After applying the above transformation, the linearized Hamiltonian \eqref{eq:orig lin Hami} turns to
	\begin{align}\label{eq:bogo lin Ham}
	\tilde{H}_\mathrm{lin}
	= \Delta_{B} \beta^\dagger \beta
	+ \omega_{c} c^\dagger c
	+ \mathcal{G} (\beta + \beta^\dagger)(c + c^\dagger), 
	\end{align}
	with the modified detuning $\Delta_{B} = \sqrt{\Delta_{b}'^2 - \lambda^2}$ and coupling $\mathcal{G} = \mu G$.
	
	\section{Dynamics}\label{sec:dynamics}
	The dynamics of this system is described within the framework of the master equation as
	\begin{align}\label{eq:org master eq}
	\dot{\rho} =-i \comm{H_\mathrm{lin}}{\rho} + \mathcal{L}_b \rho + \mathcal{L}_c \rho,
	\end{align}
	where $\rho$ is the density operator, and the Liouvillian superoperator $\mathcal{L}_j$ represents the dissipation of the system mode $j \in \{b, c\}$ from the environment.
	In the Markov and rotating wave approximations, it reads
	\begin{align}
	\mathcal{L}_j \rho = \frac{\kappa_j}{2} \qty{(\bar{n}_j +1) \mathcal{D} [j] \rho
	+ \bar{n}_j \mathcal{D}[j^\dagger] \rho },
	\end{align}
	where $\bar{n}_j = (\exp[\hbar \omega_j / k_B T] -1 )^{-1}$ is the mean thermal occupation of the bosonic mode $j$ and $\mathcal{D}$ is the Lindblad superoperator defined by
	\begin{align}
	\mathcal{D} [o] \rho = 2 o \rho o^\dagger - o^\dagger o \rho - \rho o^\dagger o.
	\end{align}
	Since the magnon mode $b$ is transformed to Bogoliubov mode $\beta$ via Eq.~\eqref{eq:bogo transf}, the master equation \eqref{eq:org master eq} becomes
	\begin{align}\label{eq:bogo master eq}
	\dot{\rho} = - i \comm{\tilde{H}_\mathrm{lin}}{\rho}
	+ \tilde{\mathcal{L}}_\beta \rho
	+ \mathcal{L}_c \rho,
	\end{align}
	with the transformed Liouvillian superoperator $\tilde{\mathcal{L}}_\beta$ for the Bogoliubov mode $\beta$ as
	\begin{align}
	\tilde{\mathcal{L}}_\beta \rho =&\ \frac{\kappa_b}{2} (\bar{n}_\beta +1) \qty(2 \beta \rho \beta^\dagger - \rho \beta^\dagger \beta - \beta^\dagger \beta \rho) \notag \\
	 & +\frac{\kappa_b}{2} \bar{n}_\beta \qty(2 \beta^\dagger \rho \beta - \rho \beta \beta^\dagger -  \beta \beta^\dagger \rho) \notag \\
	 & -\frac{\kappa_b}{2} \qty[
	 m (2 \beta \rho \beta - \rho \beta^2 - \beta^2 \rho) + \mathrm{h.c.}
	 ],
	\end{align} 
	where the redefined thermal occupation of mode $\beta$ and the squeezing parameter are respectively as
	\begin{align}\label{eq:NandM}
	\bar{n}_\beta &= \bar{n}_b \cosh 2r + \sinh^2 r, \notag \\
	m &= (2 \bar{n}_b + 1 ) e^{-i(\theta + 2\phi)} \sinh r \cosh r.
	\end{align}

	By putting the linearized Hamiltonian \eqref{eq:bogo lin Ham} into the master equation \eqref{eq:bogo master eq}, the expectation values ${\expval{\bullet}} \equiv \tr(\bullet \rho)$ of the second-order moments obey the following set of coupled differential equations:
	\begin{align}
	\dot{\expval{\beta^\dagger \beta}} =&\ \kappa_b \qty(\bar{n}_\beta - \expval{\beta^\dagger \beta}) + i \mathcal{G} \qty(\expval{\beta c} + \expval{\beta c^\dagger} - \mathrm{h.c.}), \notag \\
	\dot{\expval{c^\dagger c}} =&\ \kappa_c \qty(\bar{n}_c - \expval{c^\dagger c} ) + i \mathcal{G} \qty(\expval{\beta c} + \expval{\beta^\dagger c} - \mathrm{h.c.}), \notag \\
	\dot{\expval{\beta c}} =&\ -\qty[i\qty(\Delta_{B} + \omega_c) + \frac{1}{2} \qty(\kappa_b + \kappa_c)] \expval{\beta c} \notag \\
	& - i \mathcal{G} \qty(\expval{\beta \beta} + \expval{\beta^\dagger \beta} + \expval{c c} + \expval{c^\dagger c} + 1), \notag \\
	\dot{\expval{\beta c^\dagger}} =&\ - \qty[i \qty(\Delta_{B}-\omega_{c}) + \frac{1}{2} \qty(\kappa_b + \kappa_c)] \expval{\beta c^\dagger} \notag \\
	& + i \mathcal{G} \qty(\expval{\beta \beta} + \expval{\beta^\dagger \beta} - \expval{c^\dagger c^\dagger} - \expval{c^\dagger c}), \notag \\
	\dot{\expval{\beta \beta}} =&\ \kappa_b m^\ast - \qty(\kappa_b+2i\Delta_{B}) \expval{\beta \beta} - 2i\mathcal{G} \qty(\expval{\beta c} + \expval{\beta c^\dagger}), \notag \\
	\dot{\expval{c c}} =&\ -\qty(\kappa_c + 2i\omega_{c}) \expval{c c} - 2i\mathcal{G} \qty(\expval{\beta c} + \expval{\beta^\dagger c}).
	\end{align}
	Now the focus is on the final steady solution, i.e., $\dot{\expval{\bullet}}_{t \rightarrow \infty}  = 0$ at $ t \rightarrow \infty $, and all two-order moments $\expval{\bullet}$ can therefore be solved numerically, including the mean final phonon occupancy $n_c \equiv \expval{c^\dagger c}_{t \rightarrow \infty}$ .
	
	To  better understand the inherent mechanism of the optimization from Kerr nonlinearity, we give the analytical form in the limit $\kappa_c \ll \omega_{c}$ and within weak coupling regime.
	The mean final phonon occupancy is approximately given by
	\begin{widetext}
	\begin{align}\label{eq:n_c}
		n_c \approx
		\frac{\kappa_c \bar{n}_c +  \mathcal{G}^2 \qty{\bar{n}_\beta (\chi_{-} +  \chi_{-}^\ast) + (\bar{n}_\beta+1) (\chi_{+} +  \chi_{+}^\ast) + \frac{\kappa_b}{2} [m\chi_{b}^\ast (\chi_{+}^\ast + \chi_{-}^\ast) + \mathrm{h.c.}]} + \frac{\mathcal{G}^4}{2} \qty[( \chi_{-} +  \chi_{-}^\ast  ) - ( \chi_{+} +  \chi_{+}^\ast  )] F_4}{\kappa_c + \mathcal{G}^2  \qty[( \chi_{-} +  \chi_{-}^\ast  ) - ( \chi_{+} +  \chi_{+}^\ast  )]} ,
	\end{align}	
	after keeping the leading terms of $\mathcal{G}$, with the $\mathcal{G}^4$-term factor in the numerator as
	\begin{align}\label{eq:F 4}
		F_4 =&\ (2 \bar{n}_\beta + 1) \qty[\frac{1}{\omega_c^2} - \frac{(\chi_{-} + \chi_{-}^\ast) + (\chi_{+} + \chi_{+}^\ast)}{\kappa_b}]
		 + \frac{(\chi_{-} + \chi_{-}^\ast) - (\chi_{+} + \chi_{+}^\ast)}{\kappa_b} \notag \\
		& + \frac{1}{\omega_c^2}\qty{m [\kappa_b \chi_{b}^\ast - i \omega_{c} (\chi_{-}^\ast - \chi_{+}^\ast)-1] + \mathrm{h.c.}}.
	\end{align}	 
	Here we introduce both susceptibilities of cavity and center-of-mass oscillation as
	\begin{align}\label{eq:susce}
		\chi_\pm \equiv \qty[ \frac{\kappa_b}{2} + i (\Delta_{B} \pm \omega_{c}) ]^{-1},  
		\quad	
		\chi_b \equiv \qty(\frac{\kappa_b}{2} + i \Delta_{B})^{-1}.
	\end{align}
	In reality, the damping rate of mechanical mode $\kappa_c$ can be negligible in the denominator of Eq.~\eqref{eq:n_c} so that this mean final phonon occupancy can be approximately rewritten as
	\begin{align}\label{eq:n_c approx}
		n_c \approx  
		\frac{\kappa_c \bar{n}_c}{ \mathcal{G}^2  \qty[( \chi_{-} +  \chi_{-}^\ast  ) - ( \chi_{+} +  \chi_{+}^\ast  )]} 
		+ \frac{ \bar{n}_\beta (\chi_{-} +  \chi_{-}^\ast) + (\bar{n}_\beta+1) (\chi_{+} +  \chi_{+}^\ast) + \frac{\kappa_b}{2} [m\chi_{b}^\ast (\chi_{+}^\ast + \chi_{-}^\ast) + \mathrm{h.c.}]}{ ( \chi_{-} +  \chi_{-}^\ast  ) - ( \chi_{+} +  \chi_{+}^\ast  )
			}
		+ \frac{\mathcal{G}^2}{2} F_4.
	\end{align}  
	\end{widetext}
	As shown in the first term, the net damping rate of the mechanical motion ${ \mathcal{G}^2  \qty[( \chi_{-} +  \chi_{-}^\ast  ) - ( \chi_{+} +  \chi_{+}^\ast  )]} $,
	resulting from the magnon-phonon exchange, leads to the mechanical cooling if sufficiently strong.
	However, the rest terms show that the vacuum and thermal fluctuations of magnon mode heat the mechanical displacement due to their back-action to mechanical mode through the cavity-enhanced coupling. 
	
	More notably, this result also indicates that the squeezing effect, characterized by the parameter $m$, can be exploited to reduce the second term denoted by $n_c^{(0)}$, but no impact on the first term.
	In physics, the magnon self-Kerr nonlinearity transfers to magnon squeezing via strong sideband driving to magnon mode, and hence reduces the magnon fluctuation to relieve the back-action heat, reflected by the second term $n_c^{(0)}$, in an appropriate phase.
	Consequently, this process results in cooling enhancement without destroying the original cooling reflected by the first term.
	
	In the following, our emphasis is laid on the second term $n_c^{(0)}$ which can be optimized due to the synergy between magnon Kerr nonlinearity in YIG sphere and two-tone strong pump driving to the magnon mode.
	
	\section{Cooling optimization}\label{sec:optimization}
	
	As mentioned above, by utilizing the Kerr-nonlinear effect in YIG, the second term in Eq.~\eqref{eq:n_c} or \eqref{eq:n_c approx}, i.e., $n_c^{(0)}$, is a key component to optimize the center-of-mass cooling.
	For comparison, we first consider the case without the nonlinear effect where $\bar{n}_\beta = \bar{n}_b$, $\Delta_{B} = \Delta_{b}'$, and $m=0$.
	Based on the susceptibilities \eqref{eq:susce}, the second term in Eq.~\eqref{eq:n_c approx} becomes
	\begin{equation}
		n_c^{(0)} = \frac{(2 \bar{n}_b + 1) (4 \Delta_{b}'^2+ 4 \omega_{c}^2 + \kappa_b^2)}{16 \Delta_{b}' \omega_{c}} - \frac{1}{2}.
	\end{equation}
	Then the detuning frequency $\Delta_{b}'$ should be tuned as
	\begin{equation}
		\Delta_{b}' = \frac{\sqrt{ \kappa_b^2 + 4 \omega_{c}^2}}{2}
	\end{equation}
	to minimize the $n_c^{(0)}$ as
	\begin{equation}
		n_c^{(0)} = \frac{(2 \bar{n}_b + 1) \sqrt{\kappa_b^2 + 4 \omega_{c}^2}}{4  \omega_{c}} - \frac{1}{2}.
	\end{equation}
	
	\begin{figure*}[htb]
	\includegraphics[width=1.8\columnwidth]{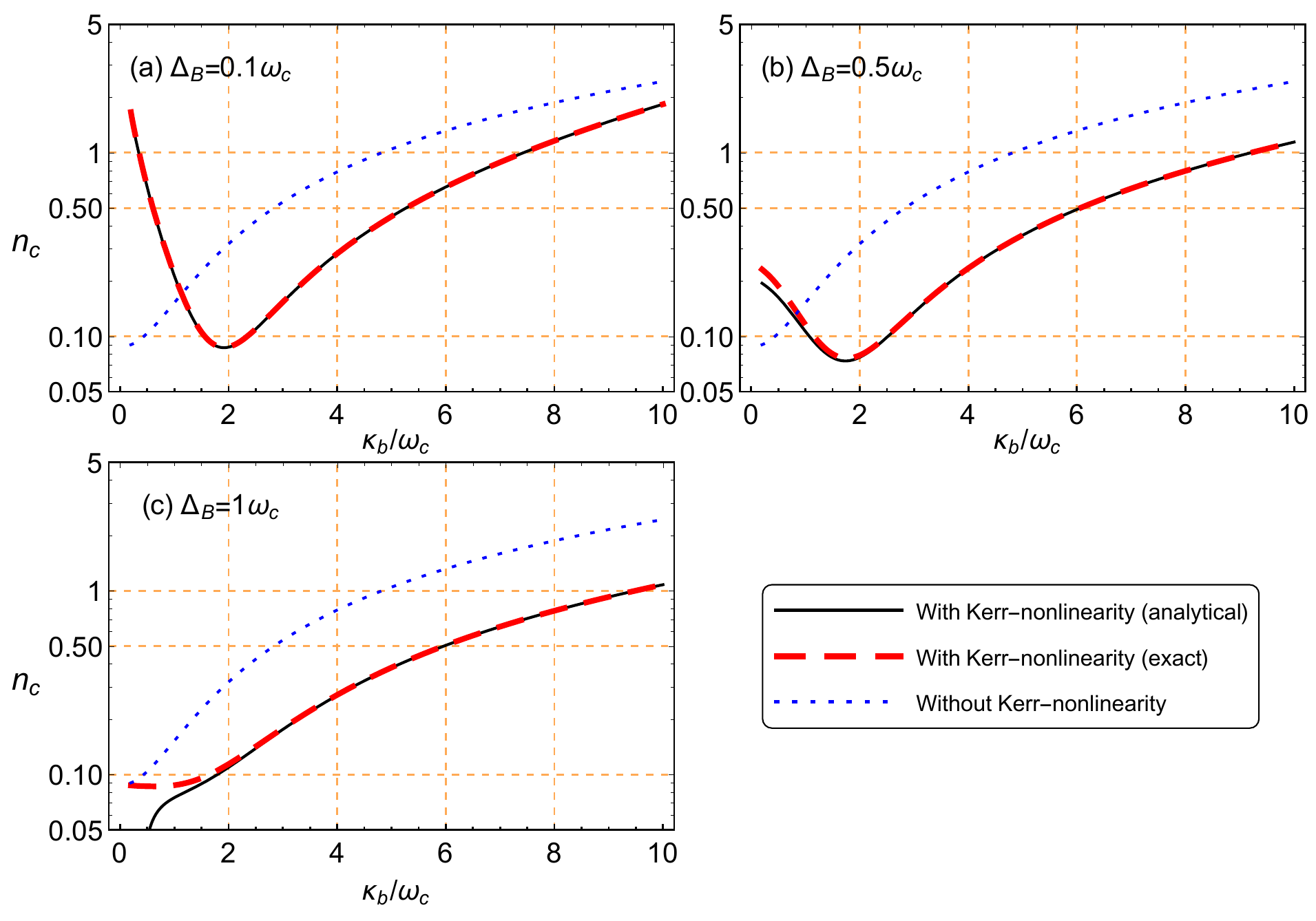}
	\caption{\label{fig:nc with kappa}
		Mean final phonon occupancy $n_c$ as a function of the magnon decay rate $\kappa_b$ for (a) $\Delta_{B} =0.1 \omega_{c}$, (b) $\Delta_{B} =0.5 \omega_{c}$, and (c) $\Delta_{B} = \omega_{c}$.
		In each panel, to display the enhancement effect from magnon self-Kerr nonlinearity, we represent both analytical form in the weak-coupling approximation (black solid line) and exact numerical result (red dashed line).
		In comparison, we also depict the exact result without the nonlinear effect (blue dotted line).
		Here the parameters are $\omega_{b} / 2\pi = 30 $~GHz, $\omega_{c}/ 2\pi = 50 $~kHz, $\kappa_c = 10^{-9} \omega_{c}$, and $\mathcal{G} (G) = \omega_{c}/5$ for the case with (without) Kerr effect, and the background temperature is $T= 0.5$ K.
		}
	\end{figure*}
	
	In our scheme, the cooling enhancement stems from Kerr nonlinearity and proper optimization.
	Utilizing the parameters $\bar{n}_\beta$ and $m$ in Eq.~\eqref{eq:NandM} with Eq.~\eqref{eq:r}, we rewrite the second term in Eq.~\eqref{eq:n_c approx} as
	\begin{widetext}
	\begin{align}
		n_c^{(0)}= \frac{2\bar{n}_b + 1}{4\Delta_{B}\qty(\chi_{-} - \chi_{+} + \chi_{-}^\ast - \chi_{+}^\ast)} \qty[ 
		\qty(\chi_{+} + \chi_{-})
		\qty(2\sqrt{\Delta_{B}^2 + \lambda^2} + \lambda \kappa_b \chi_{b} e^{i(\theta + 2 \phi)}) + \mathrm{h.c.}
		] - \frac{1}{2}.
	\end{align}
	This optimization is outlined as two steps.
	In the first place, two phases $\theta$ and $\phi$ are together chosen for matching a condition as
	\begin{equation}\label{eq:opt phase}
		\theta + 2 \phi + \arg \chi_b = \pi,
	\end{equation}
	to optimize this term as
	\begin{align}
		n_c^{(0)}= \frac{\qty(2\bar{n}_b + 1) \qty(\chi_{+} + \chi_{-} + \chi_{+}^\ast + \chi_{-}^\ast)}{4\Delta_{B}\qty(\chi_{-} - \chi_{+} + \chi_{-}^\ast - \chi_{+}^\ast)} 
		\qty(2\sqrt{\Delta_{B}^2 + \lambda^2} - \lambda \kappa_b \abs{\chi_{b}})
		 - \frac{1}{2}.
	\end{align}
	\end{widetext}
	The second step is made via adopting the condition
	\begin{equation}\label{eq:opt squeeze}
		\lambda = \frac{\abs{\chi_{b}} \Delta_{B} \kappa_b}{\sqrt{4 - \abs{\chi_{b}}^2 \kappa_b^2}}
		= \frac{\kappa_b}{2},
	\end{equation}
	where we have used the explicit form of the susceptibility $\chi_b$ in Eq.~\eqref{eq:susce} to obtain the second equality.
	Eventually, the optimal $n_c^{(0)}$ is given by
	\begin{equation}\label{eq:n_c at last}
		n_{c, \mathrm{opt}}^{(0)} = \frac{(2 n_b +1) \qty(\kappa_b^2 + 4 \Delta_B^2 + 4 \omega_c^2)}{8 \omega_{c} \sqrt{\kappa_b^2 + 4 \Delta_B^2}} - \frac{1}{2}.
	\end{equation}
	where we have used the explicit form of the susceptibility $\chi_\pm$ in Eq.~\eqref{eq:susce}.
	In addition, both conditions \eqref{eq:opt phase} and \eqref{eq:opt squeeze} should be substituted into the first and third terms in Eq.~\eqref{eq:n_c approx}.
	Even though the optimization procedure with these conditions only concentrates on the term $n_c^{(0)}$, the mean thermal occupation $n_c$ including all three terms in Eq.~\eqref{eq:n_c} still considerably reduces, which means cooling enhancement, in comparison to the case without exploiting magnon Kerr nonlinearity.
	Below, we display this optimization using the raw parameters listed in Table~\ref{tab:parameters} (Appendix~\ref{appd:para}), which are also adopted for the calculations in Sec.~\ref{sec:power}.
	
	We display the mean final phonon occupancy $n_c$ as a function of magnon decay rate $\kappa_b$ in Fig.~\ref{fig:nc with kappa}, comparing both cases with and without exploiting Kerr-nonlinearity.
	Besides, since this optimization conditions and results are acquired approximately from two approximate conditions, i.e., much small $\kappa_c$ and week coupling $\mathcal{G}$, we also exhibit both the exact numerical result and the approximate analytical form, corresponding to red-dashed and black-solid lines in Fig.~\ref{fig:nc with kappa}, to show how close they are.
	The range of magnon decay rate $\kappa_b$ chosen for this simulation is close
	to recent experimental state-of-the-art techniques  \cite{Zhang2014} and can be even smaller for ultrapure YIG \cite{Kani22}.	
	While the magnon frequency is adopted as $\omega_{b} / 2\pi = 30 $~GHz via tuning the external magnetic field, the parameters of center-of-mass displacement are frequency $\omega_{c}/ 2\pi = 50 $~kHz and damping rate $\kappa_c = 10^{-9} \omega_{c}$ \cite{Kani22} that does not outperform the recent proposal \cite{Chang10}.
	To get better cooling, we set the system with the coupling rate as $\mathcal{G} = \omega_{c}/5 = 10 $~kHz, as same as the coupling rate $G$ without Kerr effect for fair comparison, under optimal condition $\lambda = \kappa_b/2$ from Eq.~\eqref{eq:opt squeeze}, and at temperature $T=0.5$~K.
	To sum up, the mean thermal magnon and phonon occupancies are thus $\bar{n}_b = 0.06$ and $\bar{n}_c = 2.08 \times 10^5$ respectively.
	The Fig.~\ref{fig:nc with kappa} shows that our approximate form in Eq.~\eqref{eq:n_c at last} conforms to the exact result within the weak-coupling and the unresolved-sideband regimes ($ \kappa_b > \omega_{c} $).
	For $\Delta_{B} = \omega_{c}$ as shown in Fig.~\ref{fig:nc with kappa}(c), however, a discrepancy occurs within small mechanical damping rate as $\kappa_b < \omega_{c}$. 
	It implies that, in this case, the week-coupling approximation leads to unreasonably large negative contribution from the $\mathcal{G}^2$-term in Eq.~\eqref{eq:n_c approx} with explicitly form Eq.~\eqref{eq:F 4} and, thus, becomes invalid.
	Nevertheless, our optimization with the weak-coupling approximation is still reasonable in unresolved-sideband regime ($ \kappa_b > \omega_{c} $).
	Remarkably, the squeezing effect that stems from the self-Kerr nonlinearity is capable of enhancing the mechanical cooling from the magnomechanical coupling.
	More specifically, comparing the panels with various detuned frequency $\Delta_{B}$, we find that a minimum  phonon occupancy can be acquired at $\kappa_b \approx 2 \omega_{c} = 0.1$~MHz when the detuned frequency in the range 
	$\Delta_{B} \in (0.1, 0.5) \omega_{c}$.
	As the detuned frequency $\Delta_{B}$ becomes larger, even though the minimum does not exist, the Kerr effect still improves the mechanical cooling within the regime	$ \kappa_b > \omega_{c} $.
	
	\begin{figure*}[htb]
		\includegraphics[width=1.8\columnwidth]{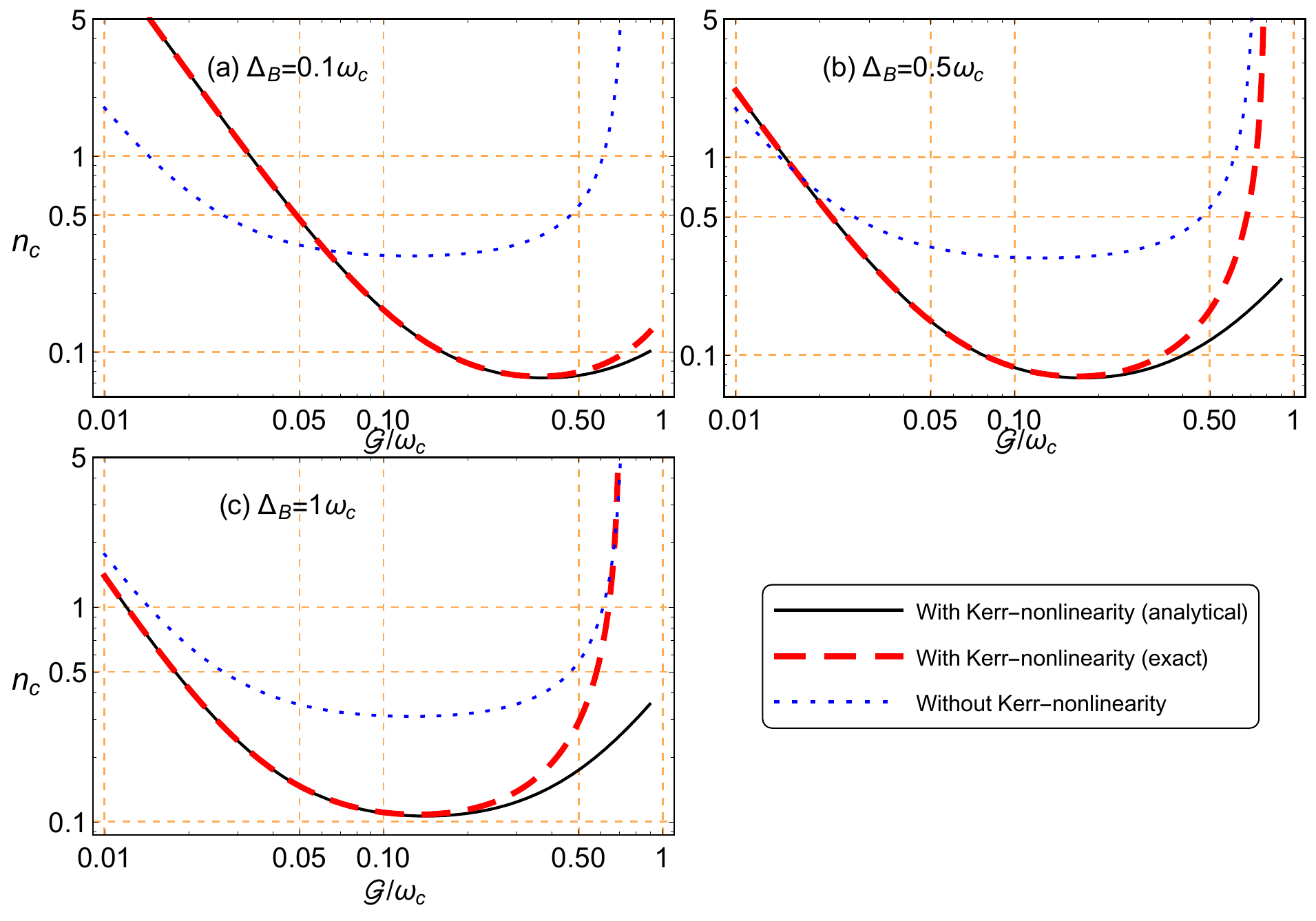}
		\caption{\label{fig:nc with G}
			Mean final phonon occupancy $n_c$ as a function of the enhanced magnon-mechanical coupling strength $\mathcal{G}$ for (a) $\Delta_{B} =0.1 \omega_{c}$, (b) $\Delta_{B} =0.5 \omega_{c}$, and (c) $\Delta_{B} = \omega_{c}$.
			Here we consider $\omega_{c} = 50$~kHz.
			In each panel, to display the enhancement effect from magnon self-Kerr nonlinearity, we represent both analytical form in the weak-coupling approximation (black solid line) and exact numerical result (red dashed line).
			In comparison, we also depict the exact result without the squeezing effect (blue dotted line).
			Here we consider $\kappa_b = 2 \omega_{c}$ and other parameters are the same as Fig.~\ref{fig:nc with kappa}.
		}
	\end{figure*}

	On the other hand, in Fig.~\ref{fig:nc with G}, we show how the cooling with squeezing enhancement depends on the coupling rate $\mathcal{G}$, in comparison to the case without exploiting Kerr-nonlinearity of magnon mode.	
	Obviously, only an appropriate coupling rate leads to optimization of cooling, and shifts with the detuned frequency $\Delta_{B}$.
	This optimization from appropriate coupling $\mathcal{G}$ can be interpreted by the trade-off relation with $\mathcal{G}$ between the first and third terms.
	It reveals that the cooling can not keep improving via injecting needless power to boost magnon-mechanical coupling since the excessively strong coupling leads to unwanted back-action heat.
	As such, if the additional power is distributed to yield sideband driving fed into magnon mode, the magnon squeezing achieves from magnon Kerr-nonlinearity and then the cooling can be improved stead.
	Besides, this figure indicates that, even in weak-coupling regime, the minimum of $n_c$, i.e., the optimal cooling, still appears and our approximate form also fits the exact result.

	Based on the relations \eqref{eq:miu} and \eqref{eq:opt phase}, the phases $\theta$ and $\phi$ can be determined.
	While the Eq.~\eqref{eq:miu} is squared to obtain two conditions as
	\begin{align}\label{eq:Re and Im}
		\mu^2 =&\ \frac{\cos(\theta + 2\phi) \qty(\sqrt{\Delta_{B}^2 + \lambda^2}\cos \theta - \lambda)}{\Delta_{B}} \notag \\ 
		&+ \sin \theta \sin(\theta + 2\phi), \notag \\ 
		0 = &- \frac{\sin(\theta + 2\phi) \qty(\sqrt{\Delta_{B}^2 + \lambda^2} \cos \theta - \lambda)}{\Delta_{B}} \notag \\ 
		&+ \sin \theta \cos(\theta + 2\phi),
	\end{align} 
	with the squeezing coefficient $r$ in Eq.~\eqref{eq:r}, the Eq.~\eqref{eq:opt phase} can be rewritten as
	\begin{align}\label{eq:the and phi}
		\sin(\theta + 2\phi) &= - \frac{2\Delta_B}{\sqrt{\kappa_b^2 + 4\Delta_{B}^2}}, \notag \\
		\cos(\theta + 2\phi) &= - \frac{\kappa_b}{\sqrt{\kappa_b^2 + 4 \Delta_{B}^2}},
	\end{align}
	via the mechanical susceptibility $\chi_b$ defined in Eq.~\eqref{eq:susce}.
	Eventually, Eqs.~\eqref{eq:Re and Im} and \eqref{eq:the and phi} can be reduced to 
	\begin{align}\label{eq:optimal theta and phi}
		\sin \theta &= -1, &\quad 
		\mu = \sqrt{\frac{\sqrt{4\Delta_{B}^2 + \kappa_b^2}}{2\Delta_{B}}}. \notag \\
		\cos 2\phi &= \frac{2\Delta_B}{\sqrt{\kappa_b^2 + 4\Delta_{B}^2}}, &\quad 
		\sin 2\phi = - \frac{\kappa_b}{\sqrt{\kappa_b^2 + 4 \Delta_{B}^2}}.
	\end{align}
	The real factor $\mu$ in our cooling enhancement mechanism is larger than unity and, according to the relation $\mathcal{G} = \mu G$, our proposal thus needs less power to drive the MW cavity if we set the same $\mathcal{G}$ and $G$ for fairly comparing both cases with and without squeezing effect.
	
	Besides, we keep our proposal in the stable regime.
	Therefore,	according to the Routh-Hurwitz criterion which contains the necessary and sufficient conditions for the stability of the system \cite{DeJesus87}, we have checked the stability of the system with the optimal parameters discussed above.
	It happens since we focus on the weak-coupling regime.
	
	\section{Power Regime}\label{sec:power}
	In this part, we present the detailed calculation for the driving powers, $\mathcal{P}_a$ and $\mathcal{P}_\pm$, which yield magnon-mechanical coupling $G$ and squeezing strength $\lambda$ in Hamiltonian \eqref{eq:orig lin Hami}.
	Eqs.~\eqref{eq:enhanced factor} manifest that the coefficients $G$ and $\lambda$ are enhanced by the factors $A_0$ and $B_\pm$, respectively.
	$A_0$ is simply given by
	\begin{equation}\label{eq:A0}
		A_0 = \frac{G}{g}.
	\end{equation}
	Without loss of generality, we can choose $B_\pm$ as
	\begin{equation}\label{eq:Bpm}
		B_- =  \sqrt{\frac{\lambda}{|k|}}, \quad
		B_+ = i\sqrt{\frac{\lambda}{|k|}},
	\end{equation}
	since the coefficient $k$ is negative in practice and the phase $\theta$ is optimized as $\sin \theta = -1$ in Eq.~\eqref{eq:optimal theta and phi}.
	The driving amplitudes $\mathcal{E}_a$ and $\mathcal{E}_\pm$ are exactly determined by these factors $A_0$ and $B_\pm$ based on the steady-state equations \eqref{eq:steady eqs}, and respectively specify the driving powers $\mathcal{P}_a$ and $\mathcal{P}_\pm$ to cavity and magnon modes.
	The parameters used below are listed in Table~\ref{tab:parameters} (Appendix~\ref{appd:para}).
	
	\subsection{Tripartite coupling and self-Kerr coefficient}\label{sec:coupling}
	Before solving Eqs.~\eqref{eq:steady eqs}, the bare coupling strengths, including tripartite coupling $g$ and self-Kerr coefficient $k$, should be obtained.
	For this purpose, we adopt the parameters mentioned in Sec.~\ref{sec:optimization} for displaying mean final phonon occupancy in Figs.~\ref{fig:nc with kappa} and \ref{fig:nc with G}.
	
	The tripartite coupling between magnon, microwave, and mechanical modes is given by \cite{Kani22}
	\begin{equation}\label{eq:tri coupling}
		g= \frac{\gamma_0}{2} \sqrt{\frac{\hbar \omega_{a} \mu_0}{V_a}} \sqrt{2 \rho_s V_s s} 
		\qty(\frac{\omega_{a}}{c})
		\sqrt{\frac{\hbar}{2 \rho_m V_s \omega_{c}}},
	\end{equation}
	with physical constants including gyromagnetic ratio $\gamma_0 = 2\pi \times 28 \ \mathrm{GHz/T}$, vacuum permeability $\mu_0 = 4\pi \times 10^{-7} \ \mathrm{N/A^2}$, YIG's ground-state spin number $s= 5/2$, and its spin density $\rho_s = 4.22 \times 10^{27} \ \mathrm{/m^3}$ and mass density $\rho_m = 5170 \ \mathrm{kg/m^3}$.  
	Moreover, we consider the typical frequencies as $\omega_{a} = \omega_{b} = 30$~GHz and $\omega_{c} = 50$~kHz of the corresponding modes defined in Eq.~\eqref{eq:org Ham}, and	the volume of cavity mode is assumed as
	\begin{align}
		V_a &= (\pi/4) L w^2 \notag \\
		&= (\pi/4)\times 0.01 \times (100 \times 10^{-6})^2
		\approx 8 \times 10^{-11} \ \mathrm{m}^3
		\notag
	\end{align}
	with cavity length $L=1$~cm and waist $w = 100 \ \mu\mathrm{m}$  \cite{Chang10}.
	Besides, the volume of the YIG sphere $V_s$ is needless in this part.	
	Accordingly, Eq.~\eqref{eq:tri coupling} yields $g/2\pi = 0.13 $~mHz.
	
	On the other hand, the self-Kerr coefficient of magnon mode is given by \cite{Shen22}
	\begin{equation}\label{eq:self Kerr}
		k = \frac{13 \hbar K_\mathrm{an} \gamma_0^2}{4 M^2 V_s},
	\end{equation} 
	with the first-order magnetocrystalline anisotropy constant $K_\mathrm{an} = - 610 \ \mathrm{J/m^3}$ and saturation magnetization  (i.e., magnetization per unit volume) $M = \hbar \gamma_0  \rho_s s$ \cite{Soykal10}
	which only involves physical constants mentioned above.
	In a macromagnet sphere with typical radius $r_s = 100 \ \mu \mathrm{m}$, Eq.~\eqref{eq:self Kerr} yields
	$k/2\pi \approx - 6.42$ nHz, 
	in accord with the previous data \cite{Shen22}.
	
	\subsection{Driving amplitudes}
	
	Combined the relations  Eqs.~\eqref{eq:A0} and \eqref{eq:Bpm} with the optimal squeezing condition \eqref{eq:opt squeeze}, we can numerically solve the steady-state equations \eqref{eq:steady eqs} to acquire the driving amplitudes $\mathcal{E}_a$ and $\mathcal{E}_\pm$ from the required coupling strengths $\mathcal{G}$ and $\lambda$ via the typical parameters.
	Apart from the parameters mentioned in Sec.~\ref{sec:coupling}, the aforementioned ones in the Figs.~\ref{fig:nc with kappa} and \ref{fig:nc with G} in Sec.~\ref{sec:optimization} are also employed in the following calculations.
	
	Superconducting microwave cavities can support mode frequency as $\omega_{a} = 30$~GHz and possess an extremely high-quality factor exceeding $Q_a \sim 10^{11}$ \cite{Kuhr2007, Romanenko2020}.
	Thus, we adopt the damping rate $\kappa_a = \omega_{a}/ Q_a \sim 10^{-1}$~Hz.
	For center-of-mass vibration, we take the mechanical frequency $\omega_{c} = 50$~kHz and damping rate $\kappa_c = 10^{-9} \omega_{c}$ \cite{Kani22} that does not outperform the recent proposal \cite{Chang10}.
	For YIG sphere, the magnon decay rate is $\kappa_b = 2\omega_c = 0.1 $~MHz, close to recent experimental state-of-the-art techniques \cite{Zhang2014}, and can even be smaller for ultrapure YIG \cite{Kani22}.
	While the reasonable magnon-mechanical coupling and detuning between magnon mode and cavity driving are $\mathcal{G} = 0.2 \omega_c = 10$~kHz and $\Delta_{B} = \omega_c $ from Fig.~\ref{fig:nc with G}, the squeezing coefficient should be taken via the optimal condition \eqref{eq:opt squeeze}, i.e., $\lambda = \kappa_b/2 = 50$~kHz.
	In addition, the detuning between magnon mode and magnon driving is considered as $\Delta_{d} = 5\omega_{c} = 250$~kHz.
	
	After substituting these parameters into Eqs~\eqref{eq:A0} and \eqref{eq:Bpm}, we obtain the steady-state values as $A_0 = 7.7 \times 10^7$ and $B_- = -i B_+ = 2.8 \times 10^6$.
	The strong driving for magnon mode is still under the assumption of  low-lying excitations, i.e., $|B_\pm|^2 \approx 7.8 \times 10^{12} \ll 10^{17}$ in this YIG size \cite{Li2018}, making our results valid.
	Then we completely solve the steady-state equations \eqref{eq:steady eqs} to acquire the driving amplitudes $\mathcal{E}_a \approx 4.5 \times 10^7$~Hz, $\mathcal{E}_+ \approx 5.9 \times 10^{11}$~Hz, and $\mathcal{E}_- \approx 8.4 \times 10^{11}$~Hz, still lower $\mathcal{E} \approx 7.1 \times 10^{14}$~Hz adopted in preview work \cite{Li2018}.
	
	\subsection{driving powers}
	Microwave cavities driven by laser with power $\mathcal{P}_a$ gain driving amplitude as
	$$|\mathcal{E}_a| = \sqrt{ \frac{\mathcal{P}_a \kappa_a}{\hbar \omega_{a}} }.$$
	To achieve $\mathcal{E}_a \approx 4.5 \times 10^7$~Hz, the power $\mathcal{P}_a \approx 0.4$~$\mu$W is required.
	For magnon mode, the driving amplitude and power are related via driving magnetic field $\mathtt{B}_\pm$ as \cite{Li2018}
	\begin{equation}
		|\mathcal{E}_\pm| = \frac{\sqrt{5}}{4} \gamma_0 \sqrt{N_s} \mathtt{B}_\pm, \quad
		\mathcal{P}_\pm = \frac{\mathtt{B}^2_\pm}{2\mu_0} A_s c,
	\end{equation}
	where involve some physical constants (including gyromagnetic ratio $\gamma_0 / 2\pi = 28 $ GHz/T, vacuum permeability $\mu_0 = 4 \pi \times 10^{-7}$ N/$\mathrm{A}^2$, and speed of light $c = 3 \times 10^8$ m/s), driving magnetic field $\mathtt{B}_\pm$, and the total number of spins $N_s = \rho_s (4/3) \pi r_s^3$ and cross-sectional area $A_s = \pi r_s^2$ with spin density $\rho_s = 4.22 \times 10^{27} \ \mathrm{m}^{-3}$ and radius $r_s = 100 $ $\mu$m. 
	The driving amplitudes $\mathcal{E}_+ \approx 5.9 \times 10^{11}$~Hz and $\mathcal{E}_- \approx 8.4 \times 10^{11}$~Hz required in this scheme thus correspond to the magnetic fields as 
	$\mathtt{B}_+ \approx 0.3$~$\mu$T and $\mathtt{B}_- = 0.4$~$\mu$T.
	Eventually, the powers for driving magnon mode are given by $\mathcal{P}_+ \approx 0.3$~$\mu$W and $\mathcal{P}_- \approx 0.6$~$\mu$W.
	
	To sum up, while the coupling strength $\mathcal{G}$ minimizing the phonon occupancy $n_c$ (shown in Fig.~\ref{fig:nc with G}) requires the power as
	$\mathcal{P}_a \approx 0.4$~$\mu$W 
	for cavity mode $a$, the optimal condition in Eq.~\eqref{eq:opt squeeze} for the cooling  improvement necessitates two-tone driving with powers
	$\mathcal{P}_+ \approx 0.3$~$\mu$W and $\mathcal{P}_- \approx 0.6$~$\mu$W
	for magnon mode $b$.
	
	\section{Conclusions}\label{sec:conclusions}
	Based on magnonic Kerr nonlinearity, we put forward a quantum squeezing scheme to enhance the center-of-mass cooling of a levitated YIG sphere trapped inside a microwave cavity.
	In this system, the cavity mode is simultaneously coupled to the magnon mode hosted in the YIG sphere and the mechanical mode of center-of-mass motion.
	We show that driving the magnon mode via	an appropriate pump field converts the magnonic nonlinearity into degenerate squeezing,
	thereby considerably improving the center-of-mass cooling compared with the previous work excluding the Kerr nonlinearity.
	We also explicitly derive an optimal condition for the squeezing rate. 
	Under the condition, effective cooling enhancement persists even in the unresolved-sideband regime.
	Moreover, we find that the cooling can only be optimized with an appropriate driving-enhanced coupling between magnon and mechanical modes.
	This result reveals that excess power for enhancing the magnon-mechanical coupling is detrimental to the cooling.
	Nevertheless, such extra power should be redirected to drive the magnon mode, generating the beneficial squeezing from the magnonic Kerr-nonlinearity.
	Eventually, after accounting for all conditions, we obtain the pump powers required for optimal cooling in a typical physical system, which fall within the sub-$\mu$W range.
	Our work provides a feasible route to achieve quantum ground-state preparation for fundamental quantum tests and quantum technological applications.

	\section*{Acknowledgement}
	This research is supported by
	the Quantum Science and Technology-National Science and Technology Major Project (Grant No. 2024ZD0301000) and
	Key R{\&}D Program of Zhejiang (Grant No.~2026C01004).
	
	\appendix
	\section{system parameters}\label{appd:para}
	Table~\ref{tab:parameters} provides a list of the system parameters used for Fig.~\ref{fig:nc with kappa}, Fig.~\ref{fig:nc with G}, and the calculations in Sec.~\ref{sec:power}.
	\begin{table*}[htbp]\label{tab:parameters}
	\centering
	\caption{List of system parameters.}
	\begin{tabular}{l|l|l}
		\hline
		\bf{Quantity} & \bf{Symbol} & \bf{Value}   \\
		\hline \hline
		gyromagnetic ratio & $\gamma_0$ & $2\pi \times 28 \ \mathrm{GHz/T}$ \\
		vacuum permeability	& $\mu_0$ &  $4\pi \times 10^{-7} \ \mathrm{N/A^2}$	\\
		spin density of YIG & $\rho_s$ & $4.22 \times 10^{27} \ \mathrm{/m^3}$ \cite{Zhang2014} \\
		mass density of YIG & $\rho_m $ & $ 5170 \ \mathrm{kg/m^3}$  \\
		ground-state spin number of YIG & $s$ & $5/2$ \\
		first-order magnetocrystalline anisotropy constant of YIG & $K_\mathrm{an} $ & $ - 610 \ \mathrm{J/m^3}$ \cite{Shen22} \\
		mode frequency of superconducting microwave-cavity  & $\omega_{a} / 2\pi $ & 30~GHz \cite{Kuhr2007} \\
		quality factor of superconducting microwave-cavity & $Q_a$ & $10^{11}$ \cite{Kuhr2007, Romanenko2020} \\
		volume of cavity mode & $V_a$ & $(\pi/4)\times 0.01 \times (100 \times 10^{-6})^2
		\approx 8 \times 10^{-11} \ \mathrm{m}^3$ \cite{Chang10} \\
		magnon mode frequency & $\omega_{b}/ 2\pi $ & 30~GHz \\
		damping rate of the magnon mode & $\kappa_b / 2\pi $ & 0.1~MHz \cite{Kani22} (agree well in order with Ref.~\cite{Zhang2014}) \\
		saturation magnetization & $M$ & $ \hbar \gamma_0  \rho_s s$ \\
		radius of YIG sphere & $r_s$ & $ 100 \ \mu \mathrm{m}$ (for volume $V_s$ and cross-sectional area $A_s$) \\
		mechanical frequency & $\omega_{c} / 2\pi $ & 50~kHz \\
		quality factor of mechanical mode & $Q_c$ & $10^9$ (does not outperform the recent proposal \cite{Chang10}) \\
		tripartite coupling & $g$ & Eq.~\eqref{eq:tri coupling} \cite{Kani22} \\
		Magnon self-Kerr coefficient & $k$ & Eq.~\eqref{eq:self Kerr} \cite{Shen22} \\
		\hline
	\end{tabular}
	\end{table*}

	\bibliography{apsreflist_cool}
\end{document}